\documentclass[conference]{IEEEtran}
\IEEEoverridecommandlockouts
\usepackage{cite}
\usepackage{amsmath,amssymb,amsfonts}
\usepackage{algorithmic}
\usepackage{graphicx}
\usepackage{textcomp}
\usepackage{layouts}

\usepackage[inline]{enumitem}
\setlist[enumerate]{label=(\roman*)}

\usepackage[dvipsnames]{xcolor}
\usepackage[binary-units=true,per-mode=symbol,group-separator={,},detect-weight=true,detect-family=true]{siunitx}
\DeclareSIUnit{\dollar}{USD}
\DeclareSIUnit{\ether}{ETH}
\DeclareSIUnit{\gas}{gas}
\usepackage{xspace}
\usepackage{tikz}
\usepackage[hidelinks]{hyperref}
\usepackage{balance} 

\usepackage{soul}

\usepackage{booktabs}
\usepackage{multirow}

\usepackage{units}

\usepackage{acro}

\definecolor{darkgrey}{RGB}{54,54,54}
\definecolor{lightgrey}{RGB}{102,102,102}

\definecolor{brown}{HTML}{a52a2a}
\definecolor{darkcyan}{HTML}{0a888a}

\newcommand{\ie}{\textit{i.e.},\xspace}
\newcommand{\eg}{\textit{e.g.},\xspace}
\newcommand{\cf}{\textit{cf.}\xspace}

\newcommand{\code}[1]{\texttt{#1}}

\usepackage{adjustbox}
\usepackage{fontawesome5}
\usepackage{booktabs}
\usepackage{wasysym}
\usepackage{tabularx} %
\usepackage{multirow}
\usepackage{arydshln} %

\newcommand{\acro}[3]{%
\DeclareAcronym{#1}{
  short = #2,
  long  = {#3},
}%
}

\acro{scm}{SCM}{supply chain management}

\usepackage{setspace}

\newif\ifauthor
\authortrue

\ifauthor
    \usepackage{tikz}
    
    \newcommand\copyrightnotice{%
    \begin{tikzpicture}[remember picture,overlay]
    \node[anchor=south,yshift=15pt] at (current page.south) {\parbox{\dimexpr\textwidth-\fboxsep-\fboxrule\relax}{\copyrighttext}};
    \end{tikzpicture}
    }
    \IEEEoverridecommandlockouts
    \IEEEpubid{\makebox[\columnwidth]{Author manuscript. \hfill} \hspace{\columnsep}\makebox[\columnwidth]{ }}
\else
    \newcommand\copyrightnotice{}
    \IEEEoverridecommandlockouts
    \IEEEpubid{\makebox[\columnwidth]{978-1-6654-9538-7/22/\$31.00~\copyright2022 IEEE \hfill} \hspace{\columnsep}\makebox[\columnwidth]{ }}
\fi

\newcommand{\name}{CCChain\xspace}

\begin{document}

\title{Scalable and Privacy-Focused Company-Centric Supply Chain Management}

\author{

\IEEEauthorblockN{Eric Wagner$^{*,\dagger}$, Roman Matzutt$^\dagger$, Jan Pennekamp$^\dagger$, Lennart Bader$^*$,\\ Irakli Bajelidze$^\dagger$, Klaus Wehrle$^{\dagger,*}$, and Martin Henze$^{\ddagger,*}$}

\IEEEauthorblockA{$^*$\textit{Cyber Analysis \& Defense}, Fraunhofer FKIE $\cdot$ \{firstname.lastname\}@fkie.fraunhofer.de\\
$^\dagger$\textit{Communication and Distributed Systems}, RWTH Aachen University $\cdot$ \{lastname\}@comsys.rwth-aachen.de\\
$^\ddagger$\textit{Security and Privacy in Industrial Cooperation}, RWTH Aachen University $\cdot$ henze@cs.rwth-aachen.de\\
}
}

\maketitle

\begin{abstract}
Blockchain technology promises to overcome trust and privacy concerns inherent to centralized information sharing.
However, current decentralized supply chain management systems do either not meet privacy and scalability requirements or require a trustworthy consortium, which is challenging for increasingly dynamic supply chains with constantly changing participants.
In this paper, we propose \name{}, a scalable and privacy-aware supply chain management system that stores all information locally to give companies complete sovereignty over who accesses their data.
Still, tamper protection of all data through a permissionless blockchain enables on-demand tracking and tracing of products as well as reliable information sharing while affording the detection of data inconsistencies.
Our evaluation confirms that \name offers superior scalability in comparison to alternatives while also enabling near real-time tracking and tracing for many, less complex products.
\end{abstract}

\begin{IEEEkeywords}
supply chain management; blockchain; permissionless; deployment; tracing and tracking; privacy
\end{IEEEkeywords}

\section{Introduction}

In 2012, the Food Safety Authority of Ireland detected significant unlabeled traces of horse meat DNA in food products~\cite{Agnolietal2016Food}.
What became known as the Horsemeat Scandal resulted in a lasting reduction of sales numbers and customer trust~\cite{Agnolietal2016Food}.
Reliable information sharing can preemptively detect issues and their underlying cause leading to such events within food and manufacturing supply chains~\cite{Gonczoletal2020Blockchain}.
Even during normal operation, such information sharing enables \eg the anticipation of bottlenecks in supply and demand~\cite{Fiala2005Information} or the sharing of product-specific information for further processing~\cite{Flynnetal2010The} and with end-customers~\cite{Maliketal2018ProductChain:}.
However, \ac{scm} systems with the capability for such reliable information sharing are still not in widespread use, mostly due to trust and privacy concerns~\cite{Gonczoletal2020Blockchain}.

Consequently, recent research efforts focus on realizing decentralized \ac{scm} systems that alleviate trust and privacy issues inherent to centralized information management~\cite{Engelenburgetal2018A}.
Lately, these efforts have concentrated on using blockchain technology as promising backbone for such decentralized \ac{scm} architectures.
The resulting proposals for \ac{scm} can be broadly grouped into two categories:
\begin{enumerate*}
	\item systems built on top of \emph{permissionless} blockchains~\cite{Tian2016An,Kimetal2018Toward,Westerkampetal2018Blockchain-Based,Wangetal2019Smart,Westerkampetal2020Tracing}, which address trust but not privacy concerns and suffer from scalability issues, and
	\item systems realized on top of \emph{permissioned} blockchains~\cite{Maliketal2018ProductChain:,Pennekampetal2020Private,Baderetal2021Blockchain-Based, Abeyratneetal2016Blockchain, Maliketal2021PrivChain:, Maliketal2021TradeChain:}, which require massive planning among collaborators to establish a trusted consortium to operate the blockchain.
\end{enumerate*}
\copyrightnotice{}%

While some prototypical or even production-ready decentralized \ac{scm} systems are already deployed, those deployments are restricted to small-scale operation with only a select few partners~\cite{Gonczoletal2020Blockchain}.
In practice, we identify two major concerns arising from these architectures.
First, large numbers of (international) collaborators
make it exponentially harder to establish a consortium trusted by collaborators and customers.
More importantly, related work mostly ignores the dynamics of entities seamlessly joining and leaving a \ac{scm} system.
This oversight is particularly limiting when \ac{scm} systems incrementally grow their user base as they gain popularity.

In this paper, we propose \name as an incrementally deployable, scalable, and privacy-aware \ac{scm} system.
\name gives each company control over its own data and by default hides data from externals.
Notably, the integrity of locally stored data is still ensured by regularly publishing a cryptographic witness over all recent events to a public ledger.
Thus, no data can be retrospectively altered.
Whenever an external entity needs access to particular information, \eg to identify the reason for a faulty product, exactly the information relevant to the product's manufacturing steps can be shared, without providing access to any additional data.

\section{Supply Chain Management Systems}
\label{sec:background}

Although information sharing along supply chains offers manifold benefits~\cite{Fiala2005Information,Flynnetal2010The,Maliketal2018ProductChain:,Pennekampetal2019Dataflow}, companies are still hesitant to engage in such efforts due to privacy and trust concerns~\cite{Wangetal2019Making,Wangetal2019Understanding,Gonczoletal2020Blockchain,Pennekampetal2019Dataflow,Pennekampetal2019Towards}.
Meanwhile, data authenticity and privacy in scalable \ac{scm} systems is still challenging~\cite{Gonczoletal2020Blockchain,Pennekampetal2020Secure},
as centralized solutions lead to information asymmetry that primarily benefits the powerful entities in the supply chain~\cite{Abeyratneetal2016Blockchain}.
Here, blockchain technology shows major potential to improve the status quo by realizing decentralized \ac{scm} that mitigates many of the concerns of centralized deployments.
In the following, we outline still open challenges of \ac{scm} that surface when either relying on permissionless or permissioned blockchains.

\vspace*{2mm}
\textbf{Permissionless Blockchain-based SCM.}
Publicly available permissionless blockchains~(\eg Ethereum~\cite{Wood2014Ethereum:}) provide a ready-to-use trusted infrastructure that offers significant advantages for \ac{scm} systems over centralized and/or dedicated alternatives.
Initially, such systems only stored integrity-protected information for individual products~\cite{Tian2016An,Kimetal2018Toward}, before several proposals tokenized not only final products but also intermediate products to enable the tracing of a product's entire manufacturing process over multiple hops~\cite{Westerkampetal2018Blockchain-Based,Wangetal2019Smart,Westerkampetal2020Tracing}.
These approaches link physical goods to a unique digital representation~(\eg via RFID chips), which is immutably stored on the blockchain.
This digital representation is then enriched with information as the product traverses the supply chain.
Although permissionless blockchains offer a high level of trust, they also suffer from scalability and privacy issues, as storing data becomes costly and all transactions are public.

\vspace*{2mm}
\textbf{Permissioned Blockchain-based SCM.}
Scalability and privacy issues of permissionless blockchains can partially be mitigated by permissioned blockchains, \ie blockchains operated by a set of authenticated users.
Utilizing such a dedicated blockchain for the information sharing along a supply chain promises faster and better scaling networks~\cite{Maliketal2018ProductChain:,Pennekampetal2020Private,Baderetal2021Blockchain-Based, Abeyratneetal2016Blockchain,Maliketal2021PrivChain:, Maliketal2021TradeChain:}.
While such fragmentation limits the interoperability across supply chains and dynamic partnerships, they promise to solve privacy and scalability issues.
In permissioned systems, privacy can be achieved by segmenting information and encrypting each part separately, \eg via attribute-based encryption~\cite{Pennekampetal2020Private,Baderetal2021Blockchain-Based}.
Combined with an entity that enforces access control~\cite{Pennekampetal2020Private,Baderetal2021Blockchain-Based,Abeyratneetal2016Blockchain}, data availability and privacy concerns can be further improved.
Still, it remains challenging for \ac{scm} systems based on permissioned blockchains to define and achieve transparency and being flexible w.r.t.\ participating companies, \ie the joining and leaving of companies.

\section{\name: A Deployable SCM System}
\label{sec:design}

Current \ac{scm} systems are either not scalable and privacy-preserving or are designed for a static setup of entities that comprise a single supply chain.
The lack of support for incremental deployability of scalable systems, which is necessary for today's dynamic economy, thus limits current systems' applicability in the real world.
Therefore, we present our novel \textbf{C}ompany-\textbf{C}entric supply \textbf{Chain} management system that addresses these limitations.
\name orients itself on designs built on permissionless blockchains while heavily restricting the amount of published data to ensure scalability and privacy.
In the following, we present \name's architecture in detail.

\subsection{Immutable and Traceable Data Records}

\begin{figure}[t]
	\centering
	\includegraphics[trim=0 40 0 0]{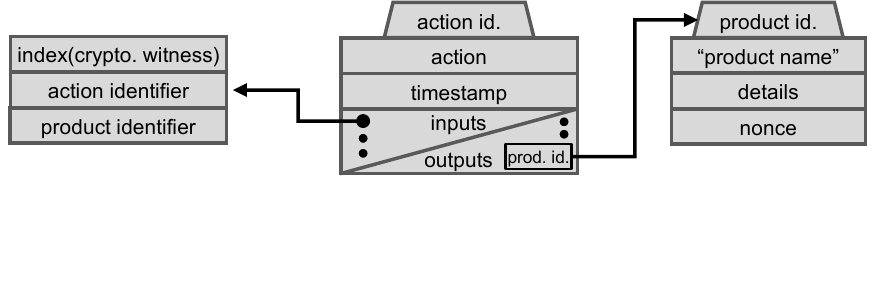}
	\caption{
		\name maintains product records with a set of actions, and links them such that fine-granular data access to individual products is possible.	
	}
\vspace*{-4mm}
	\label{fig:data_structure}
\end{figure}

We first describe how \name represents the supply chain's production graph with data records that are stored locally by the entity responsible for the respective production step.
Data records represent actions that describe transformations or ownership transfers of the referenced products.
Hence, the gradual transformation and transfer of raw materials to the final product via actions form the final production graph.
As we show in Figure~\ref{fig:data_structure}, every action consists of a unique \emph{action identifier} (a hash value computed over all its fields), an \emph{action type}, and a \emph{timestamp}.
Additionally, actions reference (potentially multiple) \emph{product identifiers} as inputs and \emph{product records} as outputs, depending on the action type.

Similarly, product records have a unique \emph{product identifier} (the hash value over all of its fields), and they contain a descriptive \emph{product name}, a list of related \emph{details} such as produced quantities, and a \emph{nonce}.
Crucially, actions only indicate the identifier of output product records to protect the confidentiality of non-relevant information when publishing an action.
Furthermore, the nonce prevents outsiders from guessing product records.
In contrast, an input product is represented by a reference to a previous output product.
Each output product can only be consumed once, which has to be respected by companies since any misbehavior is retrospectively detectable during product tracing or third-party audits.

In total, \name supports six action types.
First, companies can \code{create} new products that do not require previous inputs, \eg gaining resources from excavation.
A company can further \code{produce} a product by consuming one or multiple input products.
Next, one company can \code{export} products, which other companies then can \code{import} to represent an ownership transfer~(cf.~Section~\ref{sec:design:transfer}).
Correspondingly, companies can \code{buy} or \code{sell} components or products to external parties, \ie trade partners that still fail to support \name.
Overall, these six action types can represent a product's full life cycle along the supply chain, enriched with supplementary collected data.

\subsection{Ensuring the Correctness of Locally Stored Information}
\label{sec:design:correctness}

\name ensures the confidentiality of their product records by only storing them locally by default.
However, \name has to ensure the integrity and consistency of product records to prevent fraud and to add value over existing \ac{scm} systems.
Thus, these records must be immutable once they are referenced by an action.
We consider potential problems stemming from generating fraudulent product records orthogonal to our work; corresponding approaches engulf, for instance, tamper-resistant sensor~\cite{Baderetal2018Smart,Pennekampetal2020Secure}, regular consistency checks~\cite{Maliketal2019TrustChain:}, or detailed (lab) analyses of random product samples~\cite{Dudderetal2017Timber}.
\name can easily integrate corresponding advancements and, conversely, also improve related auditing processes:
For example, through the management of access rights to a company's product records, auditors can gain retroactive data access such that potential misbehavior is always detectable.

To ensure that locally stored data records cannot be altered, companies are required to periodically (\eg once a day) upload a \emph{cryptographic witness} covering all recent products and actions to a permissionless blockchain.
These witnesses consist of the Merkle roots obtained from the Merkle tree~\cite{Merkle1987A} over all actions generated by one company within the last period.
Only after a witness is published, \ie being immutable, can other companies rely on its information.
The existence of an action is then validated via exchanged Merkle proofs.

\subsection{Transferring Products Across Company Borders}
\label{sec:design:transfer}

The \emph{export} and \emph{import} actions are unique as they connect the production graphs across two companies.
These actions always occur in pairs~(\cf~Section~\ref{sec:design:transfer}), yet, fraudulent and incorrect data must always be attributable to a single entity.

Ideally, companies create and share their export actions right before shipping goods to the receiving company and prove they have been recorded on the blockchain.
This way, the receiving company learns a verifiable and retroactively publishable claim that the goods have been shipped, and it can create and publish the corresponding import action when receiving the goods.
Hence, both trading companies agree that the shipment has concluded successfully and that the respective actions are properly persisted for later audits.
However, this protocol enables malicious companies to misbehave, motivating a fallback protocol to resolve potential disputes.

If the exporting company receives no corresponding import action within a reasonable time, it files a complaint on the blockchain.
The complaint includes identifiers of the blinded importing company, the affected products, and an indicator of whether the export action should be retracted entirely, \eg in case of problems with the shipping process, or converted into a sell action because goods were received but not registered in \name.
The blinding is based on a secret previously established between both companies to protect the accused company's privacy during the dispute resolution.

After recognizing a complaint, the accused company has two options.
First, it can choose not to respond, which we interpret as an agreement to the accusation and to reverting the previous export operation.
In this case, the exporting company can overwrite the export action by consuming the corresponding product identifiers twice.
Second, the accused company can publish a \emph{Merkle proof} showing that the company previously published an import action that matches the accusing company's export action, \ie the complaint was unwarranted.
If a product is shipped without a published or transmitted export action, the importing company can employ a similar protocol:
It first announces the (blinded) company identifier and the relevant product identifiers.
Without any response, it uses a buy action instead of an import action even if both parties use \name.
If further questioned, the company can refer to the unanswered complaint on the blockchain.
Otherwise, the exporting company publishes the proof of the export action to resolve the dispute.
Consequently, both companies involved in a product transfer always own a shareable proof (for future use) about the existence of the counterpart of their respective import or export action.

\begin{figure}[t]
	\centering
	\includegraphics[trim=0 17 0 0, width=\columnwidth]{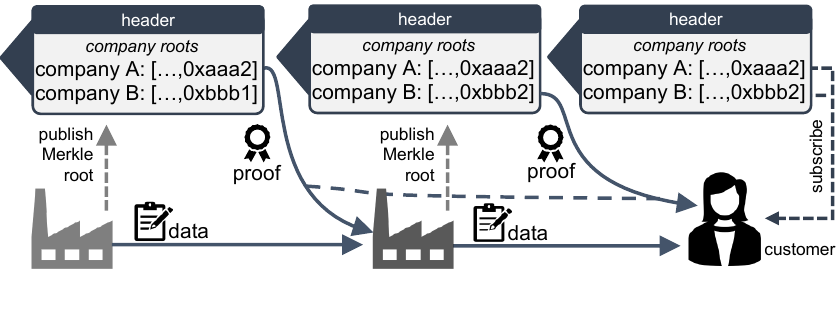}
	\caption{In \name, the basic tracing and tracking operation works by iteratively requesting product-related information from its creator. This process ensures companies keep the sovereignty of their data.}
	\vspace*{-6mm}
	\label{fig:tracing}
\end{figure}

\subsection{Tracing and Tracking Product Lifecycles with \name}

As companies and end-customers increasingly care about products' sustainable, ethical, and conflict-free sourcing and manufacturing~\cite{Gonzalez-Benitoetal2016Role,De-Angelisetal2017The,Crossetal2010Rough}, a critical feature of modern \ac{scm} systems is the ability to perform multi-hop tracking and tracing~\cite{Changetal2020When}.
The tracking and tracing functionality in \name is an interactive process, where each company can decide on a case-by-case basis if it wants to share the requested information.
While information can be technically withheld, without a valid reason such an approach damages a company's reputation.
Similarly, companies can be required by law to share their data with regulators, for which the legal frameworks are currently extended, by \eg the German \emph{Act on Corporate Due Diligence in Supply Chains} from 2021~\cite{Bundestag2021Act}.

Figure~\ref{fig:tracing} illustrates the data retrieval process for an end-customer tracing a product in \name.
The customer first has to retrieve the relevant companies' Merkle roots from the public blockchain.
Then, a product's tracing information can be requested in an iterative manner, starting from the entity the product was brought from.
The company from the supply chain can answer these requests with product-related information and the Merkle proof verifying its authenticity.
By iteratively tracing a product back to its raw materials, its sources and entire lifecycle can thus be verified, while the protocol presented in Section~\ref{sec:design:transfer} ensures that the correct supply chain predecessors are shared with the customer.

While the need for entities to remain responsive could be considered a drawback, we argue that \name's dynamic data access significantly outweighs this aspect, as companies never lose control over their data.
In the end, we believe that this additional control over one's data is crucial for any \ac{scm} system to be accepted by a large community.

\section{Performance Evaluation}
\label{sec:eval}

We have introduced \name as an incrementally deployable \acl{scm} system by design and discussed its security guarantees.
In the following, we underline \name{}'s excellent performance and scalability.

\textbf{Experimental Setup.}
Our Go-based prototypical implementation covers all relevant entities and relies on MongoDB as a database.
We further realize the event-based smart contract in Solidity and interface via geth with a local Ethereum instance.
We run our evaluation on a single server (Intel Xeon Silver 4116 CPU, 196 GB of RAM) by deploying all entities in individual docker containers, with each container having access to a single core.

\subsection{\name{}'s Low Local Processing and Storage Overheads}

Before discussing the performance on a global scale, we report on the low processing and storage overhead for a company that utilizes \name{}.
Looking at \name's performance in terms of processed actions per second, we observe that the \code{create} and \code{buy} actions can be handled at an average rate of \num{79} and \num{61} thousand actions per second, respectively.
The \code{sell} and \code{produce} actions, on the other hand, require a backward reference to the input products but still achieve a more than sufficient throughput of \num{1.6} and \num{1.4} thousand actions per second.
As expected, the \code{export} and \code{import} actions achieve the least throughput at \num{240} actions per second, each.
This reduced performance follows from the need to generate and verify Merkle trees, as well as the required interactions between two \name instances.
Assuming an equal distribution of action types, with each action only carrying \SI{200}{\byte} of data, \name achieves a per-company throughput of over \SI{35}{\mega\bit\per\second}.
Consequently, even our prototypical \name client demonstrates excellent performance on a company level.

\subsection{Operational Cost of a \name{} Deployment}

Besides the local \name client, additional costs arise from \name's interaction with a public blockchain.
On March~3, 2022, the recommended gas costs~\cite{ETHGasStation2017ETHGasStation} for a Merkle root upload thus equals \SI{0.0011}{\ether} or \SI{3.18}{\dollar}.
For larger businesses, the resulting annual price constitutes \SI{1160.7}{\dollar}.
Consequently, these funds are manageable even with high transaction costs, as currently observed in Ethereum.
Alternatively, multiple companies could also register as a group and combine their individual Merkle roots, thereby sharing the cost of a single upload.
With a threshold cryptography scheme for account management, such an approach does not impair their individual privacy.
Overall, the fair operational costs of \name are easily manageable by most businesses, making its design optimal to achieve widespread real-world adoption.

\subsection{ Scalability Analysis of \name{}}

\begin{figure}[t]
	\centering
	\includegraphics[]{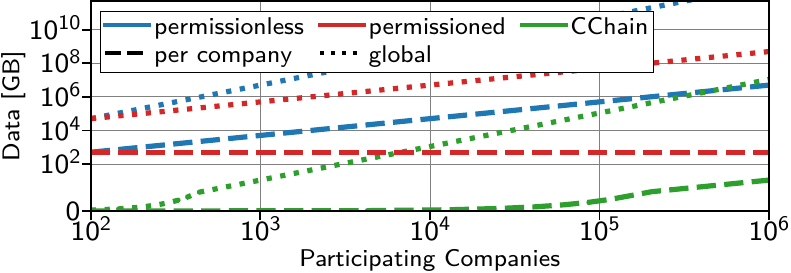}
	\vspace{-2mm}
	\caption{%
			\name is even more efficient in terms of data storage requirements than SCM systems built on top of private blockchains.		
	}
	\vspace{-6mm}
	\label{fig:scalability}
\end{figure}

To look at \name's scalability of an entire supply chain,
we now compare \name to abstract \ac{scm} systems that either operate fully on a permissionless or on a permissioned blockchain.
In the permissionless setting, we assume that all information is stored on a public ledger and thus a copy of all data must be stored by each entity.
In the permissioned setting, we expect that entities only store a copy of the blockchains corresponding to supply chains they are directly involved in.
In our context, we make a conservative estimate that each company is part of supply chains with \num{100} collaborators.

For comparability, we assume that each entity using the network stores one \SI{200}{\byte} action per second, which corresponds to \SI{5.9}{\giga\byte} of annual data per entity.
Figure~\ref{fig:scalability} illustrates how the compulsory inclusion of this data in the different decentralized \ac{scm} architectures affects the required storage as the number of participating companies grows.
On a per-company level, we notice that a permissioned architecture adds constant overhead to each company, given that the companies only store data from their (multi-hop) collaborators.
Even worse, permissionless architectures handle exponentially growing amounts of data as every company's data is stored.
In contrast, \name introduces little overhead, and only in scenarios with vast numbers of participating companies, the overhead grows as the Merkle Roots of all participants must be stored.
Thus, \name's good scalability even allows smaller businesses to join without fearing a high financial burden.

\subsection{Tracking and Tracing Performance in \name{}}

\begin{figure}[t]
	\centering
	\includegraphics[]{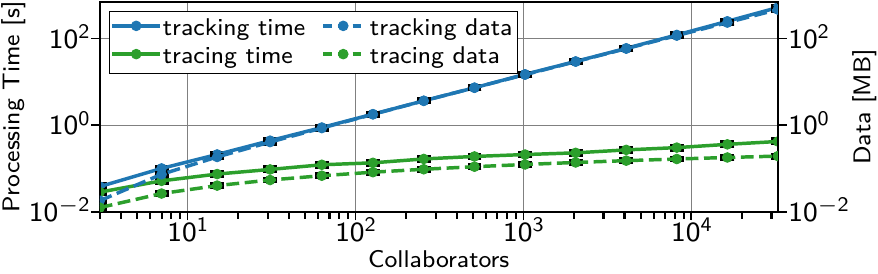}
	\vspace{-6mm}
	\caption{%
		\name allows the quick tracking and tracing of even complex products composed over thousands of manufacturing steps.
	}
	\label{fig:tracing_eval}
	\vspace{-6mm}
\end{figure}

Finally, we look at the tracking and tracing performance (\cf~\cite{Baderetal2021Blockchain-Based}).
For these measurements, we consider a binary production graph where at each non-leaf node, a new (intermediate) product is manufactured by a unique company and by consuming two previous products.
In Figure~\ref{fig:tracing_eval}, we report on the time and network traffic to track and trace products for an increasing number of entities.
We refrain from applying additional network latency as many of the requests can be parallelized.
As expected, tracing produces more overhead as it requires the revelation of the entire production graph, whereas tracking only reveals one path from one node up to the final product.
We observe a similar trend in the amount of exchanged data for the requests, information sharing, and authenticity proofs.
In our setting, \name thus traces products with \num{64} or less manufacturing steps in less than \SI{1}{\second}, while exchanging less than \SI{1}{\mega\byte} of data.
Hence, \name could even be used to request verified provenance information in real-time while shopping for food in a supermarket.

\section{Conclusion}

Information sharing along the supply chain tackles the anticipation of issues and answers the customers' and regulators' demands for proof of the sustainable, ethical, and conflict-free sourcing and manufacturing of products.
Blockchain technology now gives rise to the opportunity of designing decentralized alternatives that mitigate trust and privacy concerns of centralized systems.
However, current decentralized \ac{scm} systems lack flexibility and scalability.
We addressed this issue by proposing \name as an alternative system:
\name supports dynamic and fine-granular data access policies, which, coupled with its scalability and incrementally deployability to support organic growth, make it an attractive alternative for today's sophisticated and ever-changing \ac{scm}.

\vspace*{2mm}
{\small
	
\textbf{Acknowledgments.}
Funded by the Deutsche Forschungsgemeinschaft (DFG, German Research Foundation) under Germany's Excellence Strategy -- EXC-2023 Internet of Production -- 390621612.}

\balance

\bibliographystyle{IEEEtran}
\bibliography{paper}

\end{document}